# Accessing HID Devices on the Web With the WebHID API
## How to play the Chrome Dino Game by Jumping With a Nintendo Joy-Con Controller in One's Pocket

Thomas Steiner
tomac@google.com
Google Germany GmbH
Hamburg, Germany

François Beaufort
fbeaufort@google.com
Google France SARL
Paris, France

## ABSTRACT
In this demonstration, we show how special hardware like Nintendo Joy-Con controllers can be made accessible from the Web through the new WebHID API. This novel technology proposal allows developers to write Web drivers in pure JavaScript that talk to Human Interface Device (HID) devices via the HID protocol. One such example of a driver has been realized in the project Joy-Con-WebHID, which allows for fun pastimes like playing the Google Chrome browser's offline dinosaur game by jumping. This works thanks to the accelerometers built into Joy-Con controllers whose signals are read out by the driver and used to control the game character in the browser. A video of the experience is available.

## 1 INTRODUCTION AND BACKGROUND
Universal Serial Bus (USB) is a communications architecture that gives a personal computer (PC) the ability to interconnect a variety of devices using a simple four-wire cable. These devices are broken into various device classes. One of these classes is the Human Interface Device (HID) class. The HID class consists primarily of devices that are used by humans to control the operation of computer systems. Typical examples of HID class devices include [5]:

(1) Keyboards and pointing devices—for example: standard mouse devices, trackballs, and joysticks.
(2) Front-panel controls—for example: knobs, switches, buttons, and sliders.
(3) Controls that might be found on devices such as telephones, VCR remote controls, games or simulation devices—for example: gloves, throttles, steering wheels, and rudder pedals.
(4) Devices that may not require human interaction but provide data in a similar format to HID class devices—for example: bar-code readers, thermometers, or voltmeters.

The HID protocol was originally developed for USB devices, but has since been implemented over many other protocols, including Bluetooth. For the context of this demonstration, we focus on gamepad devices of the category (1) that connect over Bluetooth.

## 2 RELATED WORK
The Gamepad specification [1] defines a low-level interface that represents gamepad devices and allows Web applications to directly act on gamepad data. Interfacing with external devices designed to control games has the potential to become large and intractable if approached in full generality. The authors of the specification explicitly chose to narrow the scope to provide a useful subset of functionality that can be widely implemented and that is broadly useful. Specifically, they chose to only support the functionality required to support gamepads. Support for gamepads requires two input types: buttons and axes. Both buttons and axes are reported as analog values. The authors deliberately excluded support for more complex devices that may also be used in gaming contexts, including those that do motion or depth sensing, video analysis, gesture recognition, *etc.* One such example are Nintendo's Joy-Con controllers that contain gyroscopes and accelerometers.

## 3 THE WEBHID API
The WebHID API [4] closes this gap by providing a way to implement device-specific logic in JavaScript. This HTTPS-only API is asynchronous by design to prevent the website UI from blocking when awaiting HID input. This is important because HID data can be received at any time, requiring a way to listen to it. HID consists of two fundamental concepts: reports and report descriptors. Reports are the data that is exchanged between a device and a software client. The report descriptor describes the format and meaning of data that the device supports. A report descriptor describes the binary format of reports supported by the device. Applications and HID devices exchange binary data through three report types:

- **Input report:** Data that is sent from the device to the application (*e.g.*, a button is pressed.)
- **Output report:** Data that is sent from the application to the device (*e.g.*, a request to turn on the keyboard backlight.)
- **Feature report:** Data that may be sent in either direction. The format is device-specific.

To open a HID connection, a `HIDDevice` object needs to be accessed. This can either happen by prompting the user to select a device by calling `navigator.hid.requestDevice()`, or by picking one from `navigator.hid.getDevices()`, which returns a list of devices the website has been granted access to previously. The `navigator.hid.requestDevice()` function takes a mandatory parameter `filter` used to match any device connected with a USB vendor identifier (`vendorId`), a USB product identifier (`productId`), a usage page value (`usagePage`), and a usage value (`usage`) that can

Thomas Steiner, François Beaufort

be obtained from the USB ID Repository[1] and the HID usage tables[2] document. Listing 1 shows how to connect to Joy-Con controllers.

```
// Feature detection to see if the API is supported.
if (!("hid" in navigator)) return;
// Filter on Nintendo Switch Joy-Cons.
const filters = [{
  vendorId: 0x057e, // Nintendo Co., Ltd
  productId: 0x2006 // Joy-Con Left
},{
  vendorId: 0x057e, // Nintendo Co., Ltd
  productId: 0x2007 // Joy-Con Right
}];
// Prompt the user to select a Joy-Con device.
const [device] = await navigator.hid.requestDevice({ filters });
```

Listing 1: Connecting to Nintendo Joy-Con controllers

Once the HID connection has been established, incoming input reports are handled by listening to `"inputreport"` events from the device. Those events contain the HID data as a `DataView` object (`data`), the HID device it belongs to (`device`), and the 8-bit report ID associated with the input report (`reportId`). Listing 2 shows how to detect button presses on Nintendo Joy-Con controllers.

```
device.addEventListener("inputreport", event => {
  const { data, device, reportId } = event;
  // Handle only the Joy-Con Right device and a specific report ID.
  if ((device.productId !== 0x2007) && (reportId !== 0x3f)) return;
  const value = data.getUint8(0);
  if (value === 0) return;
  const buttons = { 1: "A", 2: "X", 4: "B", 8: "Y" };
  console.log(`Pressed button ${buttons[value]}.`);
});
```

Listing 2: Listening to Joy-Con controller input reports

## 4 SECURITY AND PRIVACY

The spec authors have designed and implemented the WebHID API using the core principles defined in Controlling Access to Powerful Web Platform Features [3], including user control, transparency, and ergonomics. The ability to use this API is primarily gated by a permission model that grants access to only a single HID device at a time. In response to a user prompt, the user must take active steps to select a particular HID device. More details about the security tradeoffs can be found in the Security and Privacy Considerations section of the WebHID spec [5]. Chrome—as the currently sole implementing user-agent—inspects the usage of each top-level collection, and if a top-level collection has a protected usage (*e.g.* generic keyboard, mouse), then a website won't be able to send and receive any reports defined in that collection. The full list of protected usages is publicly available.[3] Security-sensitive HID devices (such as FIDO HID authentication devices) are blocked in Chrome.

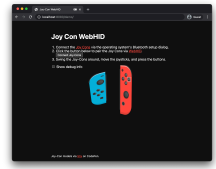
(a) Demo of the driver

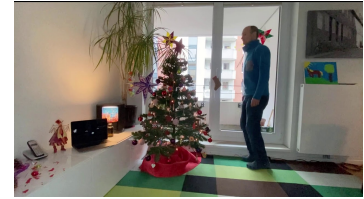
(b) Playing by jumping

Figure 1: Joy-Con WebHID and Chrome Dino WebHID

## 5 DEMONSTRATION

We have implemented a WebHID driver for Nintendo Joy-Con controllers that allows for full access to all of the controllers' buttons, axes, motion actuators, accelerometers, and gyroscopes. The project is available on GitHub.[4] Figure 1a shows the driver's demo, where the orientation of the controllers in the hands of the user is reflected virtually on the screen. To demonstrate the practicability of the driver, we have adapted the Chrome offline dino game [2] so that it can be played by jumping with a Joy-Con controller in one's pocket. This game is likewise available on GitHub.[5] A video of the experience can additionally be seen on YouTube.[6] Figure 1b shows one of the authors interact with the demonstration by jumping.

## 6 CONCLUSION

The Web platform already supports input from many HID devices. Keyboards, pointing devices, and gamepads are all typically implemented using the HID protocol. However, this support relies on the operating system's HID drivers that transform HID input into high-level input APIs. Devices that are not supported by the host's HID drivers are inaccessible to Web pages. Providing access to HID devices through the Web platform reduces installation requirements, particularly for devices that are currently only supported through host-specific applications. With our demonstration, we have motivated the existence of the WebHID API and proven its practicability.

## REFERENCES
[1] Steve Agoston, James Hollyer, and Matt Reynolds. 2020. *Gamepad*. Working Draft 29 October 2020. W3C. https://www.w3.org/TR/gamepad/.
[2] Google Blog. 2018. As the Chrome dino runs, we caught up with the Googlers who built it. (2018). https://www.blog.google/products/chrome/chrome-dino/.
[3] Dominick Ng and Rory McClelland. 2019. *Controlling Access to Powerful Web Platform Features*. Technical Report. https://goo.gle/access-to-powerful-features.
[4] Matt Reynolds. 2021. *WebHID API*. Draft Community Group Report 08 January 2021. WICG. https://wicg.github.io/webhid/.
[5] USB Implementers' Forum. 2001. *Device Class Definition for Human Interface Devices (HID), Firmware Specification—5/27/01, Version 1.11*. Technical Report. Universal Serial Bus (USB). https://usb.org/sites/default/files/hid1_11.pdf.

---

[1] http://www.linux-usb.org/usb-ids.html
[2] https://usb.org/document-library/hid-usage-tables-12
[3] https://source.chromium.org/chromium/chromium/src/+/master:services/device/public/cpp/hid/hid_usage_and_page.cc
[4] https://github.com/tomayac/joy-con-webhid
[5] https://github.com/tomayac/chrome-dino-webhid
[6] https://www.youtube.com/watch?v=HuhQXXgDnCQ